\documentclass[conference]{IEEEtran}
\IEEEoverridecommandlockouts
\usepackage{cite}
\usepackage{amsmath,amssymb,amsfonts}
\usepackage{algorithmic}
\usepackage{graphicx}
\usepackage{textcomp}
\usepackage{xcolor}
\usepackage{float}
\usepackage{subcaption}
\def\BibTeX{{\rm B\kern-.05em{\sc i\kern-.025em b}\kern-.08em
    T\kern-.1667em\lower.7ex\hbox{E}\kern-.125emX}}
\begin{document}

\title{Reservation Based Smart Parking Management 
\thanks{Identify applicable funding agency here. If none, delete this.}
}

\author{
\IEEEauthorblockN{1\textsuperscript{st} Giacomo Cabri}
\IEEEauthorblockA{\textit{dept. Physics, Informatics and Mathematics} \\
\textit{University of Modena and Reggio Emilia}\\
Modena, Italy \\
giacomo.cabri@unimore.it}
\and
\IEEEauthorblockN{2\textsuperscript{nd} Manuela Montangero}
\IEEEauthorblockA{\textit{dept. Physics, Informatics and Mathematics} \\
\textit{University of Modena and Reggio Emilia}\\
Modena, Italy \\
manuela.montangero@unimore.it
}
\and
\IEEEauthorblockN{3\textsuperscript{rd} Filippo Muzzini}
\IEEEauthorblockA{\textit{dept. Physics, Informatics and Mathematics} \\
\textit{University of Modena and Reggio Emilia}\\
Modena, Italy \\
filippo.muzzini@unimore.it
}\and
\IEEEauthorblockN{4\textsuperscript{th} Roberto Wang}
\IEEEauthorblockA{\textit{dept. Physics, Informatics and Mathematics} \\
\textit{University of Modena and Reggio Emilia}\\
Modena, Italy \\
rrobertowang@gmail.com
}
}

\maketitle

\begin{abstract}
In the framework of Smart Cities and Intelligent Transportation Systems (ITS), efficient parking management is essential to reduce urban congestion and emissions. However, current reservation-based systems often encounter a scenario in which users find their reserved slot occupied by a previous occupant who failed to vacate on time (``No PARK'' situation). 

This paper introduces a dual-mechanism architecture designed to enhance system reliability. A Reservation Module uses a dynamic size buffer of non-reservable slots to grant parking availability. A reputation-based Reward System exploits a ``star-based'' metric to incentivize punctual departures through financial penalties and access restrictions. 


The simulations conducted with the SUMO urban simulator are promising, showing that the dynamic buffer strategy provides a better tradeoff between parking availability and reservation success. By progressively adapting to users behavior, the proposed system  mitigates ``NO PARK'' instances and improves resource utilization, significantly enhancing urban viability.
\end{abstract}


\section{Introduction}
In the context of contemporary urban development, the concept of the \emph{Smart City} has become essential for managing the growing complexities of metropolitan infrastructure. Within this framework, Intelligent Transportation Systems (ITS) play a pivotal role, with smart parking identified as a primary area for optimization to improve overall urban viability~\cite{aljohani2021survey,xiao2023parking}. 

Efficient parking management is not merely a convenience but a necessity; studies have shown that optimized allocation can significantly reduce traffic congestion, lower vehicle emissions, and mitigate urban inefficiencies by maximizing space utilization~\cite{makarova2025optimizing}.

The scientific community has extensively investigated smart parking through various lenses. Early approaches, such as the ASAP system (Agent-Assisted Smart Auction-Based Parking System)~\cite{rizvi2019asap}, used agent-assisted auction mechanisms to improve resource allocation via IoT-enabled sensors. This was further evolved by the MaPark framework\cite{rizvi2020mapark}, which adopted distributed multi-agent systems to enhance robustness and adaptability in dynamic environments. Furthermore, research has addressed the need for inclusivity by designing architectures that support both equipped and non-equipped vehicles (i.e., vehicles having, or not, the ability to communicate with city infrastructure and/or other vehicles), ensuring accessibility in heterogeneous urban settings. Other paradigms have even explored infrastructure-less cooperative guidance, demonstrating that decentralized coordination among vehicles can provide effective results even without fixed sensor networks~\cite{de2025evaluating}. 

However, a persistent obstacle in reservation-based parking is the lack of guarantee regarding slot availability at the time of arrival. Specifically, a user may reserve a spot and find it occupied by a previous user who overstayed his/her estimated parking time. This discrepancy between reserved and actual occupancy creates a ``NO PARK'' scenario, which undermines the system reliability and the user trust.

To overcome these challenges, this paper proposes a dual-mechanism system focused on smart reservation management. The first component is a \emph{Reservation Module} that exploits a ``buffer'' of non-reservable slots to ensure that arriving users can be accommodated even if their specific reserved spot has not been released yet. To the best of our knowledge, this work represents the first attempt to utilize a "buffer" mechanism to address this specific challenge.
We adopt a dynamic policy to adapt the buffer size based on observed user behavior and specific area needs, such as the high variability in parking durations and user behavior observed in parking lots.
The second component is a \emph{Reward System} designed to evaluate and influence user behavior. By employing a reputation-based metric (expressed in Stars), the system rewards users who leave on time and penalizes those who overstay, using financial incentives and access restrictions to encourage compliance.
 Through simulations conducted using the SUMO urban simulator, we show that a dynamic buffer strategy provides a better tradeoff between actual parking availability and reservation success.

Even if the presented work is still preliminary, it shows the effectiveness of adopting technologies to manage parking slots in a smart way and it paves the way for future research directions.

The remainder of this paper is organized as follows.
After reporting related work  in Section~\ref{sec:relatedwork}, we present our approach in 
Section~\ref{sec:proposal}.
We introduce the experimental set up made to evaluate the proposal in Section~\ref{sec:experiments}, and we present and discuss the obtained results in Section~\ref{sec:results}. 
Finally, we draw conclusions and we discuss future work in Section~\ref{sec:conclusions}.

\section{Related Work}
\label{sec:relatedwork}

Smart parking has been widely investigated as a key component of intelligent transportation systems and smart cities, with particular attention to scalability, efficiency, and user-centric allocation mechanisms. Early approaches have explored the integration of Internet of Things (IoT) infrastructures with intelligent decision-making techniques to dynamically manage parking resources. In this context, several studies have proposed intelligent parking reservation systems with different approaches and objectives. In~\cite{hashimoto2013auction}, the authors introduced a smart parking reservation system for hybrid and electric vehicles, based on auctioning, dynamic pricing, and electricity trading. Similarly,~\cite{kokolaki2014trading} proposed a centralized, auction-based system in which drivers submit bids for public parking spaces, and a central authority coordinates the assignments and payments.

Optimization-based approaches have also been explored. In~\cite{doulamis2013improving}, an intelligent parking reservation system is proposed using interval scheduling principles: parking requests are modeled as time intervals, and the scheduler decides whether to accept a request and assign a spot or reject it. The off-street parking system in~\cite{yan2011smartparking} enables on-the-fly spot reservations, providing drivers with real-time access to available parking.

Other solutions focus on multi-channel reservation methods, allowing users to book parking via phone calls or SMS~\cite{trusiewicz2013parking}, web applications~\cite{tejas2020application}, or smartphone apps~\cite{safitri2020mobile}. A different approach is presented in~\cite{alqazzaz2018secsps}, where the system guarantees availability of parking slots by restricting access to drivers without a reservation.

Several studies also address reservation reliability and optimization. The system in~\cite{somani2018cross} requires drivers to pay a portion of the parking fee to confirm the reservation and prevent false bookings. In~\cite{mejri2016reservation}, the parking slot assignment problem is formulated as a multi-objective integer linear program, optimizing for multiple criteria, including walking distance to the destination, vehicle proximity, and parking congestion impact. Another system, described in~\cite{xiao2021collaborative}, relies on continuous updates and acknowledgments between the server and the vehicle, eliminating the need for driver interaction. Finally, the iParker system~\cite{kotb2016iparker} extends previous solutions by offering features such as flexible reservation times, and dynamic revenue and pricing models.

The authors in~\cite{rizvi2019asap} introduced ASAP, an agent-assisted auction-based parking system that leverages IoT-enabled sensors and software agents to allocate parking spaces through market-inspired mechanisms. Their approach demonstrated how auction strategies can improve resource utilization while reducing search time, highlighting the potential of agent-based coordination in urban environments.

Building on this line of research, the subsequent MaPark system proposed in~\cite{rizvi2020mapark} further formalized a multi-agent auction framework for smart parking in IoT ecosystems. Compared to earlier centralized solutions, MaPark adopts distributed agents that negotiate parking assignments, improving adaptability and robustness in dynamic urban scenarios. These contributions collectively show that auction-based and multi-agent paradigms can effectively address congestion and inefficiencies associated with traditional parking management systems.

After those works, in~\cite{aljohani2021survey}, smart parking strategies was look through the lens of market-based allocation of goods and services.

More recent studies have shifted focus toward inclusivity and urban sustainability. The work~\cite{muzzini2022smart} proposed a smart parking architecture explicitly designed to support both equipped and non-equipped vehicles, addressing a critical limitation of many IoT-based systems that assume pervasive connectivity. By combining infrastructure support with adaptable guidance mechanisms, their approach enhances accessibility and real-world deployability in heterogeneous smart city environments. In a later study~\cite{muzzini2023improving}, the authors investigated the broader impact of smart parking solutions on urban viability, showing how optimized parking allocation can reduce traffic congestion and overall urban inefficiencies through improved space utilization.

Complementary to infrastructure-based solutions, recent research has also explored cooperative and infrastructure-less paradigms. In particular, in~\cite{de2025evaluating} evaluated cooperative parking guidance using agent-based simulation without relying on fixed infrastructure. Their findings suggest that decentralized cooperation among vehicles can provide effective guidance even in partially connected environments, increasing system resilience and reducing deployment costs.

Our proposal differs from existing reservation systems by explicitly accounting for situations in which a reservation cannot be fulfilled because previous vehicles have not vacated the spot in time. We address this issue to reduce the likelihood that a reserved parking spot becomes inaccessible to the driver.

\section{Our Proposal}
\label{sec:proposal}


In this paper, we consider a scenario in which all vehicles are equipped (i.e., are able to communicate with city infrastructure) and join the proposed system,   
which has been designed to address the problem of finding parking spots in parking areas. 


We assume that access to parking areas is controlled and limited to authorized users only, by means of bars or pilomats, and that parking areas are located in different parts of the city area. 

To access a parking area, users have to place a reservation in which they also have to state for how long they will need the parking slot.  

When the user arrives at the parking area, the system recognizes that the user has a reserved spot and allows the vehicle to enter. The user can then park in any free parking spot, no specific spot is assigned to the vehicle. 
Users may extend their stay as long as this does not conflict with incoming reservations and does not affect buffer sposts.






The proposed system is based on two mechanisms. The first mechanism, called \textit{Reservation Module}, acts when a user wishes to reserve a slot in a parking area at a given time.
The second mechanism, called \textit{Reward System}, implements a review system for user's behavior, rewarding the well behaved users, and punishing the bad behaved ones.

\subsection{Reservation Module}
\label{sec:reservation}
The \textit{Reservation module} task is to ensure that incoming users will find their reserved spot free. Observe that, as the access to parking areas is restricted, only vehicles with a reservation can enter and can occupy parking spots. However, authorized users might not release their spots when supposed to.

To overcome this issue,  the \textit{Reservation module} maintains a buffer of free parking slots in each parking area. Such slots are not reservable by new incoming requests, but can be allocated to incoming users whose reserved spot has not made free yet.  



\begin{figure}
    \centering
    \includegraphics[width=\linewidth]{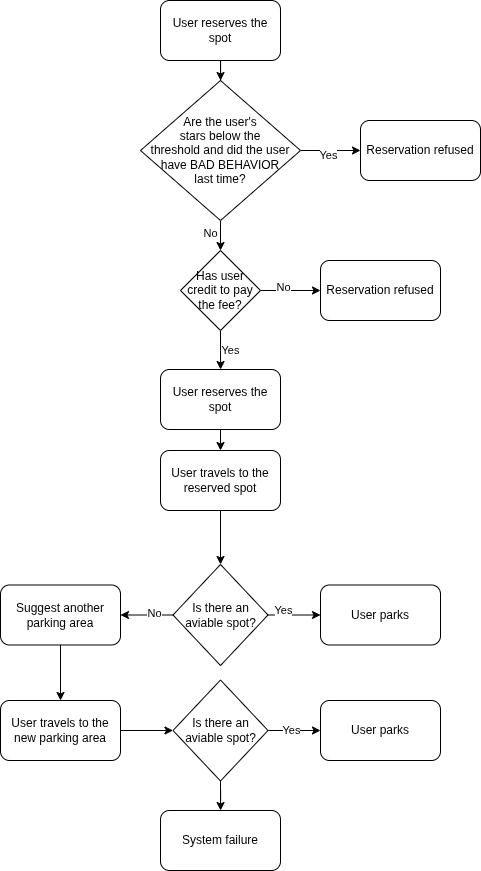}
    \caption{User reservation flow.}
    \label{fig:reservation}
\end{figure}

The reservation flow is described in Figure~\ref{fig:reservation}.
The user makes a reservation request for a spot in a parking area. The system accepts the request if the user has a good reputation (see Section~\ref{sec:reward}) and has credit to pay. So a parking spot will be available at the user's arrival time. When the user reaches the selected parking area, if a parking spot is available, the vehicle is granted access to the area and parks in any free spot. On the contrary, if the parking area does not have available spots, the system suggests another parking area close by with available parking spots and automatically places a reservation for that user in this new parking area. The user will then travel to this suggested parking area and park in any available spot, if any. If there is no parking spot available in the second area either, then the system has failed, and the user can not park with the current reservation. 


The choice to suggest only a single alternative parking area is intended to reflect realistic urban scenarios. Parking areas are usually scattered throughout the city, and users typically wish to reserve a spot close to their destination.
Recommending yet another alternative area might be impractical, as this could likely be too far from the destination of the user. Thus, the choice of where to park next is left to the user. More structured suggestion strategies can be designed (i.e., suggest more than one alternative if there are several close by parking areas) once the effectiveness of this simple one is assessed and is left to future work. 

The critical aspect of the Reservation Module is clearly how to set the buffer size: a too large buffer may lead to rejecting feasible reservation requests, whereas a too small buffer may prevent already accepted reservations from being fulfilled.


In general, the buffer size can be set in different ways: statically or dynamically. 
With a static number (or percentage) of buffer parking slots, the buffer size will be the same at all times. This is the simpler choice, but needs a careful tuning phase to choose the best suitable value based on the city (and parking area) situation.
With a dynamic buffer size, the number of parking slots in the buffer can be defined on the basis of the actual parking area situation. For example, different times of the day can lead to different buffer sizes. 

In our proposal, the dynamic policy works as follows: whenever a vehicle leaves a parking area after the end of its reservation, the system increases the buffer size associated with that area by one unit. The buffer size is initialized to a fixed constant and is increased up to a fixed maximum. Periodically, the buffer size is reset to the lowest value. In this way, the buffer is progressively adapted according to the observed behavior of users parking in that specific location. The underlying rationale is that certain areas are more prone to delayed departures, making reservation durations difficult to estimate a priori. For instance, hospital parking areas often exhibit high variability in parking time, as users may be unable to predict in advance the exact length of their stay.


Moreover, the Module also adjusts the buffer size according to the reputation of the users, as explained in the following section: 
the greater the number of low-reputation users occupying the area, the larger the required buffer size, and vice-versa. 


\subsection{Reward System}
\label{sec:reward}
The Reward System is dedicated to evaluating the behavior of the users.
The rationale is to punish users who do not respect the time of a reserved spot and to reward users who respect the reservation time, and this, hopefully, on al long run,  will educate users to ask for the most realistic reservation times possible. 

The reputation of a user is represented as a rating of \textit{Stars} on a scale from 0 to 5, and is known to the user. A higher number of stars indicates a better reputation.

The user's reputation directly influences parking fees, providing a financial incentive for the timely vacancy of the reserved slot. Furthermore, a reputation score falling below a given \textit{Stars} threshold, coupled with a recent violation, triggers a temporary suspension. Specifically, users who overstay
while their reputation is under this threshold are barred from utilizing the system-managed parking areas for one subsequent session.



If a user leaves a reserved slot in time, the system will register a \textit{Benefit} to the user. When the user collects $B$ Benefits, the system increases the \textit{Stars} of the user by one. On the contrary, when the user overstays, the system registers a \textit{Warning} for the user and resets the \textit{Benefit} counter. If the user collects $W$ \textit{Warnings}, the system decreases the number of \textit{Stars} associated with the user by one. 



\begin{figure*}[tb]
    \centering
    \includegraphics[width=0.6\linewidth]{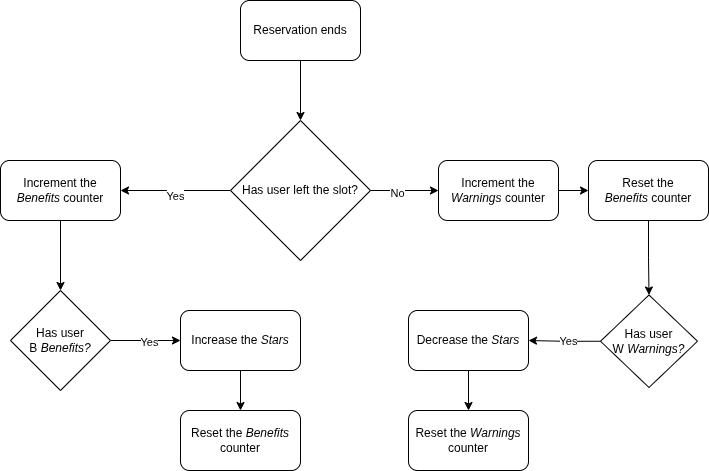}
    \caption{\textit{Reward System} flow}
    \label{fig:reward}
\end{figure*}

The complete flow of the \textit{Reward System} is reported in Figure~\ref{fig:reward}.


\section{Experimental set up}
\label{sec:experiments}

We conducted preliminary experiments to assess the effectiveness of our proposal.
We tested the proposed system using SUMO Urban Simulator\footnote{https://sumo.dlr.de/docs/index.html}. 

The scenario consists of a map of 80 reservable parking spots divided into 8 parking areas of equal size. The population is composed of vehicles with good behavior and vehicles with bad behavior. The former will always leave the parking spot in compliance with their reservation, while the latter will stay in the parking spot for an additional time of 1h. All users start the simulation with 0 \textit{Warnings}, 0 \textit{Benefits}, and 3 \textit{Stars}. 
One simulation lasts about 63 hours, in which each vehicle performs 10 parks in the area where the proposed system is in operation, and the duration of each park is randomly selected in the range of [1, 3]h. There are also two stops in the area not covered by the system, and the duration of these is always chosen at random from the interval [8, 16] hours. 
Every time a vehicle needs to park, it uses the proposed reservation system and behaves according to its good or bad behavior. 

We are interested in measuring how often the two following situations arise: 
\begin{itemize}
    \item "No Park": a vehicle with a reservation is not able to find a free parking spot, neither in the selected nor in the alternative suggested parking area. 
    \item "No Reservation": a vehicle is refused a reservation for the selected parking area. 
\end{itemize}

The first situation occurs when parking spots are still occupied by vehicles that did not depart when expected, the second when the system acknowledges that the parking area will be full at the time requested in the reservation.  

We set up two experiments: (1) we vary the number of vehicles in the population by testing the system in critical situations, i.e., starting with a population size that is equal to the number of parking spots (80), and increasing (up tp 110, step 10); (2) we vary the percentage of vehicles showing a bad behavior, from a 25\% to a 50\% of the total population size, where the size randomly varying in the range [80,110].  


We compared the proposed system (referred to as \textit{Dynamic buffer} in the following) with three alternatives: 
\begin{itemize}
    \item {\em Base line:} a reservation system with no buffer (called \textit{Static - 0 spot buffer}). 
    \item A static buffer with 10\% of total area spots in the buffer, 1 spot in our setup (called \textit{Static - 1 spot buffer}); 
    \item A static buffer with 30\% of total area spots in the buffer, 3 spots in our setup (called \textit{Static - 3 spots buffer}).
\end{itemize}

We performed 5 different runs for each configuration and report the average results.



\section{Results and Discussion}
\label{sec:results}

\begin{figure*}
    \centering
    
    \begin{subfigure}{0.45\textwidth}
        \centering
        \includegraphics[width=\textwidth]{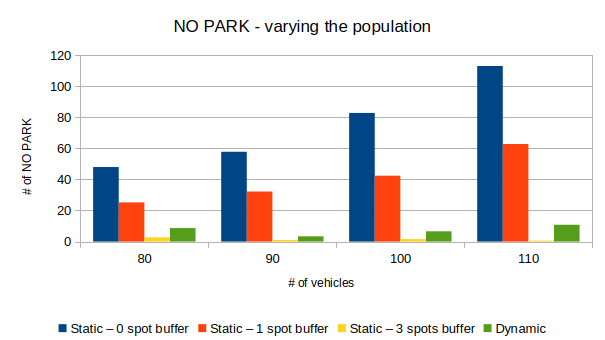}
        \caption{NO PARK counts}
        \label{fig:population-no-park}
    \end{subfigure}
    \hfill
    \begin{subfigure}{0.45\textwidth}
        \centering
        \includegraphics[width=\textwidth]{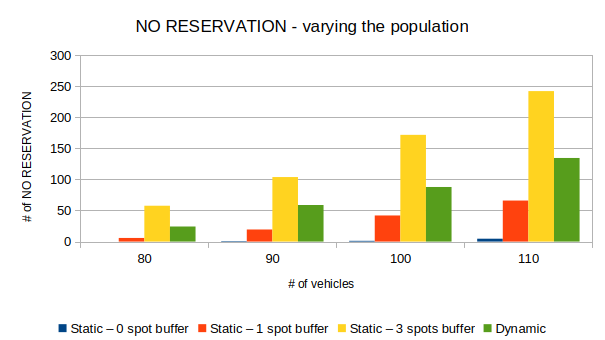}
        \caption{NO RESERVATION counts}
        \label{fig:population-no-reservatio}
    \end{subfigure}
    
    \caption{Experimental results when varying the population size.   }
    \label{fig:population}
\end{figure*}
\begin{figure*}
    \centering
    
    \begin{subfigure}{0.45\textwidth}
        \centering
        \includegraphics[width=\textwidth]{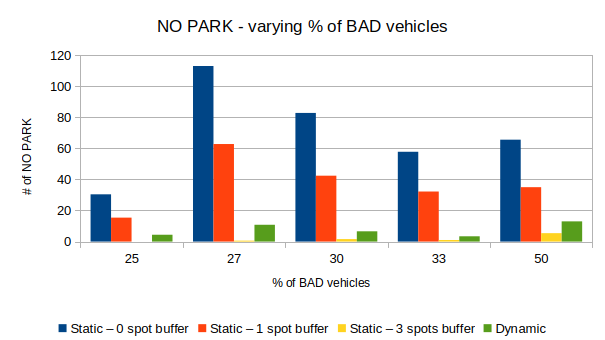}
        \caption{NO PARK counts}
        \label{fig:bad-no-park}
    \end{subfigure}
    \hfill
    \begin{subfigure}{0.45\textwidth}
        \centering
        \includegraphics[width=\textwidth]{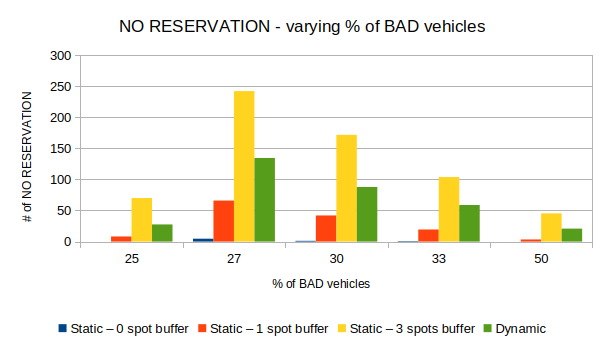}
        \caption{NO RESERVATION counts}
        \label{fig:bad-no-reservation}
    \end{subfigure}
    
    \caption{Experimental results varying the percentage of vehicles with bad behavior}
    \label{fig:bad}
\end{figure*}

Results for experiments (1) and (2) are shown in Figure~\ref{fig:population} and in Figure~\ref{fig:bad}, respectively. 

We start by observing that any use of a buffer outperforms the baseline when no buffer is used. In the baseline, a negligible number of NO RESERVATION are registered, but the largest number of NO PARK is counted. This means that users often travel to the parking area with a valid reservation but are unable to actually park, and this is the worst scenario for the user. Indeed, it is preferable for users to be informed in advance that parking is not available in a given area, rather than reaching the destination and finding no available spots.

Looking at the static buffer strategies, we can see that the smaller the size of the buffer, the larger the NO PARK counts, and the smaller the NO RESERVATION counts. This means that smaller buffers increase the number of successful reservations, as there are more parking spots made reservable. However, this comes at the cost of a higher number of failed parks, since users who overstay their allocated time prevent subsequent reservations from being honored.
On the other hand, larger buffers reduce the number of failed parks by introducing a safety margin between consecutive reservations. Nevertheless, this also leads to a reduction in the number of successful reservations, as there are fewer parking spaces available.
In particular, the {\em Static - 3 spot buffer} fails a park on very few occasions, at the expenses however, of the highest rejected reservations.  

These results suggest that no fixed buffer size is able to simultaneously minimize failed parks and maximize the number of successful reservations across all scenarios, indicating that static buffer strategies are inherently limited, as they cannot adapt to changes in user behavior.

In this panorama, the {\em Dynamic buffer} is the strategy that has the better trade-off between failed parks and rejected reservations, providing a more balanced performance and improving user satisfaction.  

It is interesting to notice that Figure~\ref{fig:bad-no-reservation} shows a decrease in the number of rejected reservations as the percentage of bad behaving vehicles increases. Since reservations are accepted or rejected solely on the basis of currently known requests, this trend cannot be solely interpreted as an improvement in the admission policy. A more plausible explanation is that longer parking durations reduce the overall turnover of vehicles, and many more remain in the parking area until the end of the simulation; therefore, the total number of reservation requests generated within the simulation horizon gets smaller. As a consequence, the absolute number of rejected requests may decrease even if the system does not become more effective. This aspect is to be better investigated in future works. 

\section{Conclusions}
\label{sec:conclusions}

In this paper we addressed the issue of managing city parking areas in a smart way, 
by proposing a parking spot reservation system enhanced with a reward system. The final aim is to design a parking policy that is able to reduce parking search times and, consequently, pollution and user frustration. 

Experimental results, obtained via the SUMO urban simulator, suggest the effectiveness of incorporating a dynamic size buffer of non-reservable slots to safeguard the reservation experience. While static buffer strategies showed a clear trade-off (where larger buffers reduced parks failures at the cost of increasing reservation rejections), the dynamic buffer strategy emerged as a better solution. By progressively adapting the number of available slots,  
the dynamic policy provides a better balance between system utility and user satisfaction. Furthermore, the \emph{Reward System} proved to be essential in fostering a more compliant user base. By exploiting a reputation metric (based on \textit{stars}), the system successfully penalizes non-compliant behavior through financial disincentives and access restrictions, thereby paving the way to increase the overall availability of spots for ``good'' users, i.e., users that respect the foreseen time to leave the parking slot. 

Despite promising preliminary results, this work is in its early stages, and several future research directions remain. 
One primary goal is to refine the dynamic buffer strategy, for example, by integrating machine learning techniques to predict peak congestion times and potential overstays with higher precision. 
A second important aspect is to improve the simulation of the behavior of users, by taking into consideration their reaction to punishments and rewards, and deeply analyze how the system reacts to changes in the user behavior to verify the increase of the availability of spots for ``good'' users. 

Expanding the simulation parameters to include more complex urban topologies and larger, more heterogeneous vehicle populations would further test the scalability and robustness of the approach.
Additionally, future work should explore the integration of this framework with real-time traffic data and multimodal transportation networks to provide a more holistic approach to urban mobility. 

\bibliographystyle{IEEEtran}
\bibliography{main}

\end{document}